**Identifying the Origin of Thermal Modulation of Exchange Bias in $MnPS_3/Fe_3GeTe_2$ van der Waals Heterostructures**

Aravind Puthirath Balan[1‡*], Aditya Kumar[1‡], Patrick Reiser[3], Joseph Vas[4], Thibaud Denneulin[4], Khoa Dang Lee[9], Tom G Saunderson[1,10], Märta Tschudin[3], Clement Pellet-Mary[3], Debarghya Dutta[3], Carolin Schrader[3], Tanja Scholz[5], Jaco Geuchies[7], Shuai Fu[7], Hai Wang[7], Alberta Bonanni[6], Bettina V. Lotsch[5], Ulrich Nowak[8], Gerhard Jakob[1], Jacob Gayles[9], Andras Kovacs[4], Rafal E. Dunin-Borkowski[4], Patrick Maletinsky[3*], Mathias Kläui[1,2*]

[1]*Institute of Physics, Johannes Gutenberg University Mainz, Staudinger Weg 7, 55128 Mainz, Germany.*

[2]*Centre for Quantum Spintronics, Department of Physics, Norwegian University of Science and Technology, 7491 Trondheim, Norway.*

[3]*Department of Physics, University of Basel, Klingelbergstrasse 82, Basel CH-4056, Switzerland*

[4]*Ernst Ruska-Centre for Microscopy and Spectroscopy with Electrons and Peter Grünberg Institute, Forschungszentrum Jülich, 52425 Jülich, Germany*

[5]*Max Planck Institute for Solid State Research, Heisenbergstraße 1, 70569 Stuttgart, Germany*

[6]*Institute of Semiconductor & Solid-State Physics, Johannes Kepler University Linz, Altenberger Straße 69, 4040 Linz, Austria*

[7]*Max Planck Institute for Polymer Research Mainz, Ackermannweg 10, 55128 Mainz, Germany*

[8]*Department of Physics, University of Konstanz, Universitaetsstrasse 10, 78464 Konstanz, Germany*

[9]*Department of Physics, University of South Florida, Tampa, FL 33620, USA.*

[10]*Peter Grünberg Institut and Institute for Advanced Simulation, Forschungszentrum Jülich and JARA, Jülich, 52425, Germany.*

‡These authors contributed equally

*Correspondence to : aravindputhirath@uni-mainz.de, patrick.maletinksy@unibas.ch, klaeui@uni-mainz.de

**Abstract**

The exchange bias phenomenon, inherent in exchange-coupled ferromagnetic and antiferromagnetic systems, has intrigued researchers for decades. Van der Waals materials, with their layered structure, provide an optimal platform for probing such physical phenomena. However, achieving a facile and effective means to manipulate exchange bias in van der Waals heterostructures remains challenging. In this study, we investigate the origin of exchange bias in $MnPS_3/Fe_3GeTe_2$ van der Waals heterostructures. Our work demonstrates a method to modulate unidirectional exchange anisotropy, achieving an unprecedented nearly 1000% variation through simple thermal cycling. Despite the compensated interfacial spin configuration of $MnPS_3$, magneto-transport measurements reveal a huge 170 mT exchange bias at 5 K, one of the largest observed in van der Waals antiferromagnet-ferromagnet interfaces. This substantial magnitude of the exchange bias is linked to an anomalous weak ferromagnetic ordering in $MnPS_3$ below 40 K. On the other hand, the tunability of exchange bias during thermal cycling is ascribed to the modified arrangement of interfacial atoms and changes in the vdW gap during field cooling. Our findings highlight a robust and easily adjustable exchange bias in van der Waals antiferromagnetic/ferromagnetic heterostructures, presenting a straightforward approach to enhance other interface-related spintronic phenomena for practical applications. A detailed study of the interface reveals migration of atoms between the layers, leading to the formation of amorphous region on either side of the van der Waals gap, underscoring the importance of precise characterization of interfaces in van der Waals heterostructures, which are often presumed to have pristine interfaces.

**Keywords:** vdW magnetic materials, exchange bias, vdW heterostructures, manganese thiophosphate, $Fe_3GeTe_2$

## 1. Introduction

Exchange bias (EB) is a unidirectional anisotropy that occurs in exchange-coupled antiferromagnetic/ferromagnetic (AFM/FM) systems, such as thin films and core-shell nanostructures. [1-3] While this phenomenon has been studied in thin-film heterostructures for over four decades, a comprehensive theoretical understanding remains a challenge for the condensed matter physics community. It is now well established, that there is not just one specific origin of the exchange bias phenomenon. Instead, several theoretical models of EB in magnetic systems have been developed, considering various possible spin configurations at the interface, including in-plane (IP) collinear, perpendicular, and out-of-plane (OOP) arrangements.[3] Furthermore, EB has been extensively studied in polycrystalline AFM/FM interfaces, with the phenomenological aspects extensively discussed by O'Grady *et al.*[4] However, most theoretical models that effectively describe EB assume well-defined AFM/FM interfaces devoid of roughness or grain boundaries.[2,5-7] In reality, thin-film heterostructures are susceptible to imperfect interfaces.[8-10] The emergence of van der Waals (vdW) magnetic materials offers a potential solution, providing a broad range of FM and AFM vdW materials. [11] Their inherent layered structure and ability to form heterostructures present a convenient means to obtain defect-free and atomically flat single-crystalline AFM/FM interfaces, renewing interest in the investigation of EB in vdW heterostructures.[12]

Although vdW AFM/FM heterostructures offer potentially ideal interfaces for EB, their weaker exchange coupling hinders their suitability for storage applications because of the presence of the vdW gap.[13] Recent efforts have focused on extrinsic control of exchange coupling by manipulating the AFM/FM interface registry, involving methods like applying pressure and laser irradiation.[14,15] Recent successful experiments have demonstrated significant EB in vdW heterostructures through coverage control and the application of gate voltage.[16,17] Additionally, the introduction of an oxide layer has shown promise, albeit challenges in controllability and repeatability.[18,19] Despite these advances, a straightforward and efficient strategy to precisely tune EB in vdW heterostructures remains elusive. Such a strategy is crucial for enabling diverse applications that require tailored EB strengths, such as magnetic recording and data storage.

Here, to understand the interplay between interface structure, magnetic order, and exchange bias, we extensively investigate $MnPS_3/Fe_3GeTe_2$ AFM/FM vdW heterostructures in a multimodal approach, combining magneto-transport anomalous Hall effect, superconducting quantum interference device (SQUID) magnetometry, scanning nitrogen-vacancy center magnetometry (S-NVM) and scanning transmission electron microscopy (STEM). Despite featuring a fully compensated AFM/FM interface [20-22], the vdW $MnPS_3/Fe_3GeTe_2$ system exhibits a remarkably high interfacial exchange-coupling strength. This gives rise to a significant EB, one of the most substantial observed in van der Waals AFM/FM interfaces so far. Notably, the strength of interfacial coupling can be modulated through thermal cycling. This thermal manipulation results in a decisive alteration in the magnitude of EB, exceeding 1000%. This thermal cycling can be implemented concurrently with the field-cooling procedure used to establish EB, thus eliminating the need for any additional steps. The unprecedented and significant alteration in EB magnitude through straightforward thermal cycling may set a platform for manipulating various other interface-related spintronic phenomena in vdW heterostructures. The subsequent section summarizes and discusses the extensive EB observed in the $MnPS_3/Fe_3GeTe_2$ system, elucidating its origin and the evolution induced by thermal cycling.

## 2. Results and Discussion

To fabricate the device, we transfer a $MnPS_3/Fe_3GeTe_2$ heterostructure onto prefabricated Au/Cr (25 nm/5 nm) Hall contacts deposited on a $Si/SiO_2$ (300 nm) substrate. We employ a Polydimethylsiloxane (PDMS) and polymethylmethacrylate (PMMA) assisted dry transfer method inside an inert atmosphere-maintained glovebox to transfer and stack flakes of $Fe_3GeTe_2$ and $MnPS_3$ in the specified order. This results in the formation of a $MnPS_3/Fe_3GeTe_2$ vdW heterostructure with a clean interface (refer to **Figure 1c** inset). To minimize the risk of oxidation due to brief exposure of the heterostructure to air during wire-bonding of the voltage and current pads for anomalous Hall measurements, we cap the obtained heterostructure with a hexagonal boron nitride (h-BN) flake. Additionally, unless otherwise stated, special care is taken to limit the air exposure time to below 30 minutes in all the devices measured during this investigation. For a detailed sample preparation procedure, please refer to the methods section. The wire-bonded device is loaded into a high-field cryostat, and anomalous Hall effect measurements are conducted using magneto-transport techniques. For comparison, we additionally carry out anomalous Hall effect measurements on freshly exfoliated $Fe_3GeTe_2$ flakes capped by h-BN and are summarised in *Supporting Information* (refer to **Figure S1**).

To induce EB, a freshly prepared device consisting of h-BN-capped $MnPS_3/Fe_3GeTe_2$ vdW heterostructure is subjected to a field-cooling process starting from 120 K (a temperature above the Neel temperature of $MnPS_3$, which is 78 K) down to the desired temperature ($T$ = 5 K) with an OOP magnetic field of +8 T & -8 T. After reaching the desired temperature, the magnetic field is ramped down to zero, and the anomalous Hall voltage ($V_{xy}$) measured by sweeping the field in the OOP direction. The normalized anomalous Hall voltage ($V_{xy}$) is used to plot the hysteresis (refer to **Figure 1a**). The EB field ($H_{EB}$) is determined using the formula $H_{EB} = \frac{(H_C^+ + H_C^-)}{2}$ where $H_C^+$ and $H_C^-$ represent the positive and negative switching fields. Surprisingly, a large negative $H_{EB}$ of magnitude 170 mT is observed at 5 K. We note that this large exchange bias is observed in a comparably thick (80 nm $Fe_3GeTe_2$) layer and while the thickness dependence of exchange bias in van der Waals heterostructures can be complex as for instance shown and discussed in reference [23], one can expect that by varying the thickness, the exchange bias can be further tuned. A complete thickness dependence study is however beyond the scope of this work. It is important to note that this large exchange bias observed is unexpected, considering the previously claimed fully compensated magnetic order of $MnPS_3$.[21-22, 24-26] Continuing our investigation, we measure the trend of $H_{EB}$ as a function of increasing temperature by executing a similar procedure at various temperatures, ranging from 5 K to 80 K (refer to **Figure 1b**). $H_{EB}$ is calculated for each measurement and plotted against the temperature, as illustrated in **Figure 1c**. It is evident that $H_{EB}$ is at a maximum (170 mT) at 5 K and gradually decreases with increasing temperature, disappearing at 40 K—well below the Néel temperature ($T_N$) of $MnPS_3$ (78 K). The presence of such a large EB that disappears at a temperature much lower than the $T_N$ indicates that the compensated antiferromagnetic order of $MnPS_3$ cannot solely explain the observed EB.

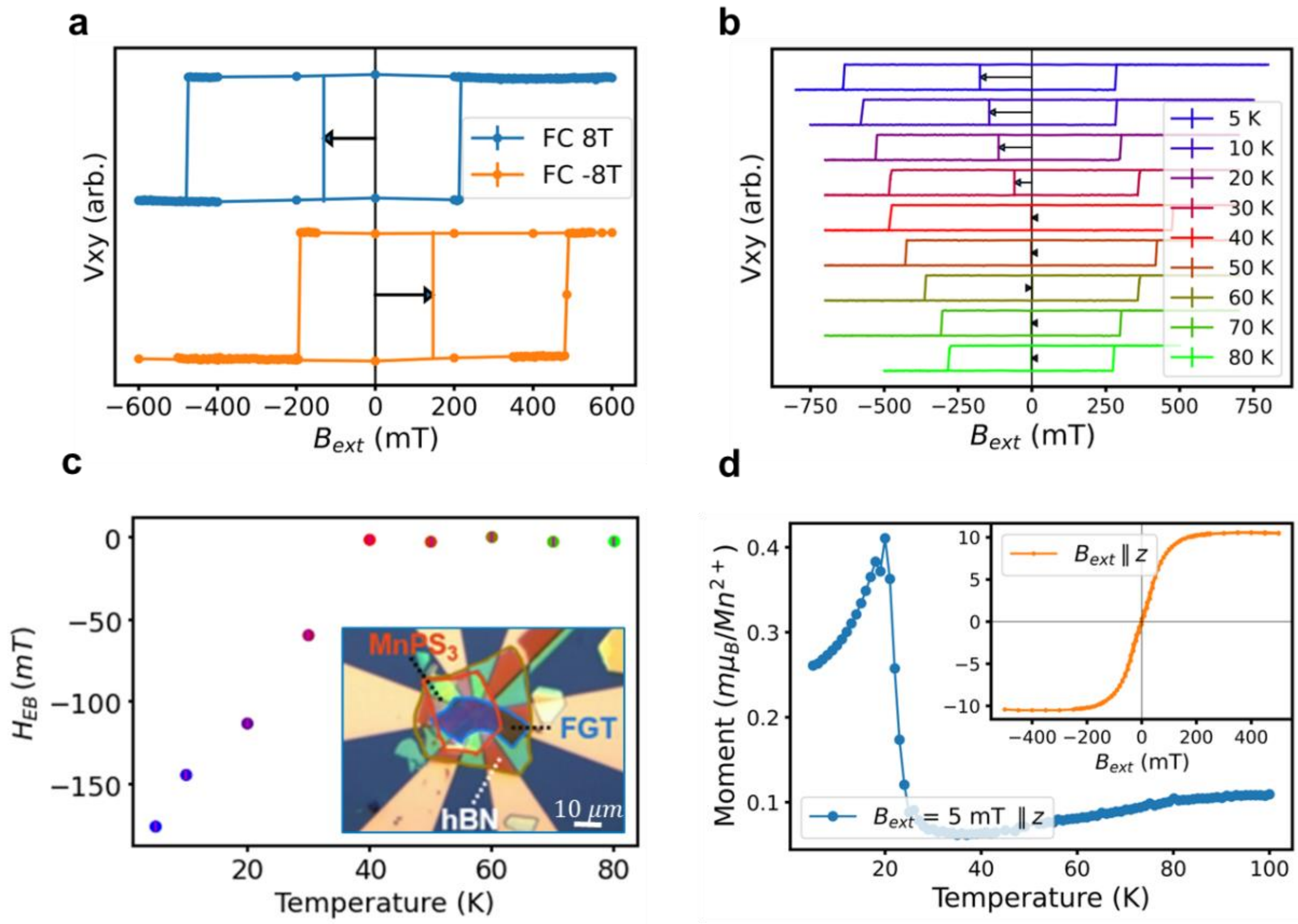

**Figure 1. a)** Anomalous Hall voltage ($V_{xy}$) as a function of the magnetic field after $\pm$ 8 T field-cooling process measured at 5 K, **b)** $V_{xy}$ as a function of the magnetic field after +8 T field-cooling and measured at temperatures ranging from 5 K to 80 K. **c)** The $H_{EB}$ $vs. T$ plot, derived from the data presented in *Figure 1b*, reveals the blocking temperature of $MnPS_3$ to be approximately 40 K. The inset provides an optical image of the measured van der Waals heterostructure. **d)** Magnetization (*M*) vs. Temperature (*T*) curve measured for the bulk $MnPS_3$ crystal in an OOP orientation. The inset displays an isothermal *M* vs. *H* curve at 5 K.

To understand more about the origin of EB, we employ dc SQUID magnetometry to obtain magnetization versus temperature (M vs. T) and magnetization versus magnetic field (*M* vs. *H*) plot at 5 K for a bulk $MnPS_3$ crystal while maintaining the magnetic field orientation OOP. The results are depicted in **Figure 1d**. Below 40 K, an anomalous magnetic moment is observed, as indicated by the onset in the magnetic moment. The emergence of this anomalous ferromagnetic moment below 40 K is also observed in the temperature dependence of spin Hall magnetoresistance in $MnPS_3$/Pt device (refer to **Figure S2** in *Supporting Information*). The onset of this anomalous moment correlates with the onset of the nonzero $H_{EB}$ (refer to **Figure 1c**). A clear saturation in M vs. *H* hysteresis at 5 K with finite coercivity is presented as further supporting evidence (see **Figure 2d** inset).

To investigate the nature and anisotropy of the anomalous ferromagnetic moment in $MnPS_3$ flakes, we next utilize S-NVM. To probe the static magnetic state of a few-layer $MnPS_3$ of comparable thickness to the $Fe_3GeTe_2$/$MnPS_3$ heterostructure under investigation, S-NVM is performed on isolated $MnPS_3$ flakes at 4.5 K. The nitrogen-vacancy (NV) center can be scanned across the sample to record the emanating stray magnetic field. This enables the inference of the underlying magnetization, with a typical spatial resolution of 30-100 nm (**Figure 2a** and details in the methods section). The examined sample comprises two isolated $MnPS_3$ flakes capped by an h-BN flake (see **Figure 2a**).

We find that a clear stray magnetic field arises from the edge of the $MnPS_3$ flake, indicating an uncompensated magnetic moment that deviates from its expected compensated antiferromagnetic order (see **Figure 2b**). A linecut is taken across the flake edge with an applied field of 150 mT along the NV axis and a magnetic model of a homogeneous magnetization is fitted to the stray field. The best agreement between the model and the data is found for Mn magnetic moments pointing along the IP direction parallel to the NV center projection to the sample plane. A magnetization of $M = 2170$ (±40) $\mathrm{m}\mu_B/\mathrm{nm}^2$ is obtained corresponding to a bulk magnetization of $15.8$ (±0.3) $m\mu_B/\mathrm{uc}$ for the flake thickness of $57.5$ (± 4.5) nm obtained by the simultaneously recorded topography (refer to **Figure 2c**). Using these parameters, the stray field of the flake is simulated, and a good qualitative agreement between data and simulation indeed confirms the easy plane anisotropy of the Mn moments. This finding is in good agreement with the recent report that suggests $MnPS_3$ undergoes a spin-reorientation transition, characterized by Mn moments that go from an OOP Heisenberg-type collinear antiferromagnetic order to the IP *XY-type* order (refer to the schematic in **Figure S2b** in the *Supporting Information*).[27] Such a low-temperature spin reorientation transition has also been independently observed by H. Han *et al.* in their recent experiments. [28] A larger magnetization (see Figure 2d) is observed for the magnetic hysteresis (M vs. H) measured with the sweeping field oriented IP with respect to the bulk $MnPS_3$ crystal as compared to that when the sweeping field is oriented OOP, which provides further supporting evidence for the IP anisotropy.

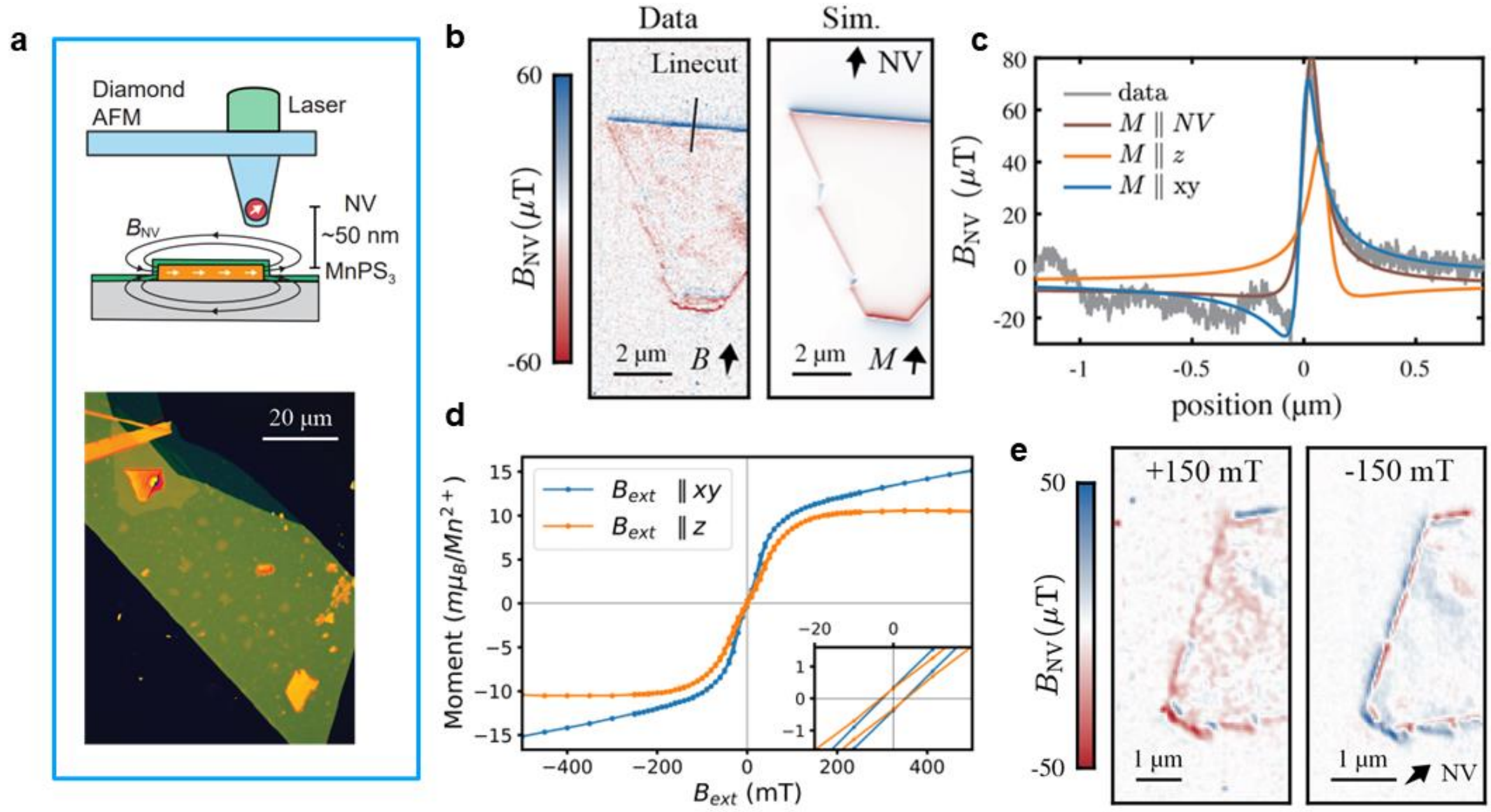

**Figure 2**. **a)** Schematics of S-NVM setup: A NV center is brought into contact with the surface, and the stray magnetic field of the sample is recorded by performing an ODMR measurement at each pixel with the microscope image of the studied sample underneath. It comprises two $MnPS_3$ flakes of ~50nm height capped by h-BN. **b)** Stray magnetic field image of the flake recorded at $B_{ext}$ = 150 mT applied along the NV axis as indicated by the arrows. Simulation of the stray magnetic field assuming IP orientation of the magnetization. **c)** Linecut taken across the edge of the flake as indicated by the line. Different magnetization models are used to fit the data. The best agreement is found with the IP orientation of the flake. **d)** DC SQUID magnetometry data of a bulk crystal taken at 5 K. Though bulk $MnPS_3$ is reported to be an ideal Heisenberg AFM, a finite magnetic moment with remanence is observed, suggesting a transition in spin-ordering at low temperatures. **e)** Stray magnetic field image for opposite field polarities. Reversing field polarity between +150 mT and –150 mT yields opposite polarity of the stray field, indicating a switchable weak ferromagnetic ordering.

In agreement with the measurements on the bulk crystal, at lower fields, the uncompensated moment shows a linear field dependence of $(\partial M/\partial B_{ext}) = 92\ (\pm 9)\ m\frac{\mu_B}{\text{Mn}}/\text{T}$ (refer to **Figure S3** in *Supporting information*) with a finite remanent magnetic moment of $1.5\ (\pm 0.7)\ \text{m}\mu_B/\text{Mn}$ close to zero-field. Importantly, the remanent moment maintains the magnetization direction unchanged at lower fields even after reversing the polarity, as evident from identical red contrast observed in images acquired at $\pm 5$ mT (refer to **Figure S4** in *Supporting information*). It is apparent from the S-NVM measurements that the remanent moment can potentially be bound to the AFM state of $MnPS_3$ and persists even above the maximum applied field of $\mp 500$ mT (maximum field reachable by the used vector magnet) along the NV direction. However, the stray fields measured at $\pm 150$ mT, as indicated in **Figure 2e**, exhibit an inverted contrast. This observation suggests that few-layer $MnPS_3$ flakes possess a larger switchable magnetic moment (below $\pm 150$ mT). This magnetic moment displays a linear variation with the applied field and predominates over the discernible remanent moment bound to the AFM state of $MnPS_3$, thereby contributing to the finite coercivity (refer to **Figure S3** in *Supporting information*).

Albeit of a soft magnetic nature, the field required to switch the uncompensated moment of the $MnPS_3$ flake is greater than 5 mT, whereas the bulk crystal has a coercive field of less than 5 mT (refer to **Figure 2d** *inset*). This contrasting behavior between bulk crystal and an exfoliated flake could be attributed to confinement effects.[20] Nevertheless, the few-layer $MnPS_3$ flake exhibits a remanent magnetic moment bound to the AFM state of $MnPS_3$. This distinctive magnetic moment serves as a pinning source, thus possibly explaining the observed EB in the $Fe_3GeTe_2/MnPS_3$ heterostructures.

The findings from bulk SQUID magnetometry and high resolution and high sensitivity S-NVM measurements unequivocally establish the presence of net uncompensated magnetic moment in $MnPS_3$. The observed EB can be ascribed to the pinning of $Fe_3GeTe_2$ by the intrinsic weak anomalous magnetic moment associated with the AFM state in $MnPS_3$ flakes at temperatures below 40 K. This phenomenon arises due to a low-temperature spin-reorientation transition from OOP Heisenberg-type order to IP *XY*-type order.[27,28] The emergence of a robust EB resulting from a low-temperature compensated antiferromagnetic (AFM) to weak ferromagnetic (FM) transition has not been previously documented in vdW heterostructures.

This observation has repercussions that could enable spintronic phenomena in $MnPS_3$ and other similar vdW systems, where such transitions can occur.

Having established the origin of EB, we next study the evolution of the $H_{EB}$ strength. We find that the observed $H_{EB}$ at a specific temperature undergoes a substantial variation, characterized by a dynamic evolution across successive iterations while maintaining experimental conditions unchanged. To systematically study this we revisit device 1 (D1), an h-BN-capped $MnPS_3/Fe_3GeTe_2$, and conduct 15 consecutive anomalous Hall voltage measurements at 5 K while implementing an 8 T field-cooling from 120 K and are depicted in **Figure 3a**. We determine $H_{EB}$ for each measurement and are plotted in a bar diagram (refer to **Figure 3b**). Surprisingly, despite maintaining identical experimental conditions, we observe different values of $H_{EB}$ after every heating and field-cooling cycle. To assess the universality of this observed metamorphosis, we fabricate two additional h-BN-capped $MnPS_3/Fe_3GeTe_2$ devices (Device 2 and Device 3) and follow a similar measurement protocol to obtain $H_{EB}$ for 15 consecutive field-cooling measurements. All three devices consistently demonstrate that a free-standing vdW AFM-FM interface is dynamic under thermal cycling, strongly influencing interface-related phenomena, particularly EB. The distribution of the evolving EB observed for 15 consecutive field-cooling measurements at 5 K for D1, D2 and D3 are summarized in *Supplementary* **Figure S5**. However, the statistical distribution of EB among these three devices are not identical as evident from *Supplementary* **Figure S6** that refers to an indeterministic variation throughout 15 cycles. To understand this intriguing thermal evolution of EB quantitatively, we consider a simplistic expression for $H_{EB}$ in an exchange-biased AFM/FM system, provided by the modified Meiklejohn–Bean model, as delineated in equation (1) presented below [29].

$$H_{EB} = \frac{J_{EB}}{\mu_0 \, M_{FM} \, t_{FM}} \quad (1)$$

where $J_{EB}$ is the interfacial exchange coupling energy, $M_{FM}$ is the saturation magnetization of the FM, and $t_{FM}$ is the thickness of the ferromagnetic layer. Now, the thermal cycling-induced evolution of EB can only be influenced by the parameter of the numerator $J_{EB}$ while the denominator remains unchanged since the materials remain unaltered. Nevertheless, the numerator $J_{EB}$, directly linked to the interface characteristics, may be contingent upon the intercrystalline distance ($\xi$)– the separation between respective crystalline regions of AFM and FM separated by a vdW gap. Recently, *Huang et al.* conducted measurements of EB in FM/AFM vdW heterostructures and demonstrated significant variations in EB with modified vdW gap induced by laser irradiation,[15] suggesting that the thermal cycling-assisted modifications to the vdW gap could potentially contribute to the observed evolution of EB.

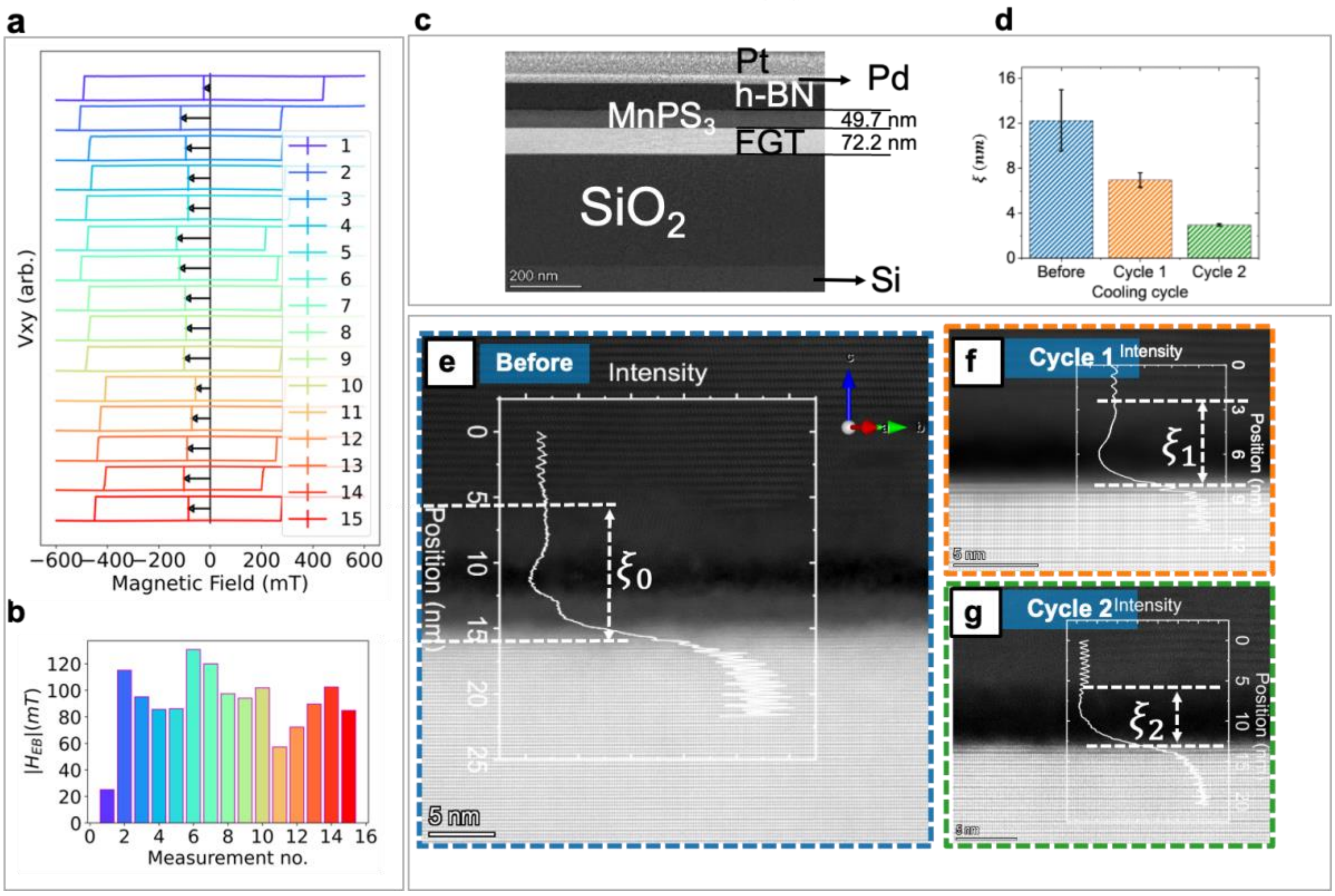

**Figure 3.** Evolution of EB due to change in van der Waal gap caused by heating-cooling cycles**. a)** AHE hysteresis loops measured after each field-cooling cycle consecutively for 15 times. **b)** The magnitude of EB varies between 25 mT and 132 mT. More measurements on additional samples can be found in the supporting information. **c)** Low magnification cross-sectional HAADF-STEM image of the $Fe_3GeTe_2$/$MnPS_3$/h-BN vdW heterostructure. **d)** A bar graph summarising the observed modification of intercrystalline distances ($\xi$) measured in the images (e, f, g) at room temperature before (e) and after two consecutive cooling and heating cycles (f &g), **e)** a high-resolution HAADF-STEM image of the sample, and $\xi_0$ is the intercrystalline distance measured at room temperature The measurements were repeated two more times after cooling-heating cycles **f)** and **g),** and $\xi_1$& $\xi_2$ are the corresponding intercrystalline distances measured**.** The intercrystalline distances measured after each cycle ($\xi_1$& $\xi_2$) differed from the initial value ($\xi_0$).

By examining $H_{EB}$ obtained during 15 consecutive cycles for device 1 (**Figure 3a** & **3b**), we can see that the value of $H_{EB}$ varies between a minimum of 25 ($\pm 5$) mT and a maximum of 132 ($\pm 5$ ) mT which corresponds to an enhancement of $> 500\ \%$ in magnitude. Although device D1 shows a larger mean $H_{EB}$ (- 95 mT) in 15 measurements compared to D2 (- 61 mT) and D3 (- 68 mT) (refer to **Figure S6** in *Supporting information*), the variation of the magnitude of $H_{EB}$ is larger ($\sim 1000\ \%$ ) in device 2 (D2) and device 3 (D3) (refer to **Figure S5** in *Supporting information*). Such a huge enhancement in $H_{EB}$ by merely thermal cycling is unprecedented and calls for further understanding of the underlying mechanism. For that, we construct a heterostructure with an h-BN-capped $MnPS_3$/$Fe_3GeTe_2$ configuration on a Si/$SiO_2$ (300 nm) substrate, and we perform cross-sectional STEM measurements during two consecutive cooling-heating cycles. The findings are presented in **Figure 3c-g**. The low-resolution high-angle annular dark-field STEM (HAADF-STEM) image of the h-BN/$MnPS_3$/$Fe_3GeTe_2$ cross-section is provided in **Figure 3c**. High-resolution STEM imaging of the vdW $MnPS_3$/$Fe_3GeTe_2$ heterostructure interface is conducted at room temperature. Refer to *Supporting Information* **Figure S7** for a summary of STEM sample fabrication and measurements conducted. The scans are repeated at multiple regions to account for any spatial non-uniformity at the interface. A low -resolution STEM image with larger field of view can be seen in **Figure S8**. By taking the mean of $\xi_0$ obtained at different regions, we obtain a value of $\xi_0 \cong 12\ (\pm 3\ )$ nm (see **Figure 3e**) which we define as the initial thickness of exchange coupled interfacial FM layer equal to the distance between the crystalline region (inter-crystalline distance) of $MnPS_3$ and $Fe_3GeTe_2$. It is evident from the HRSTEM image that $\xi$ is constituted of the vdW gap and an amorphous region formed by intermigration of atoms between $MnPS_3$ and $Fe_3GeTe_2$ as evident from the STEM-EDX elemental mapping (refer to **Figure S9** in *Supporting information*). Subsequently, we reevaluate the interface twice at room temperature, each time on a freshly prepared lamella from the same stack following the ex-situ cooling of the heterostructure to liquid nitrogen temperature and subsequent warming to room temperature (for experimental details, please refer to the respective methods section). Remarkably, $\xi$ undergoes significant modification, as evident from the corresponding cross-sectional STEM images (see **Figure 3f** *&* **3g**). The values obtained for $\xi$ after two thermal cycles are $\xi_1 \cong\ 7\ (\pm 1\ )$ and $\xi_2 \cong\ 3\ (\pm 0.03)$ nm, respectively, and are summarised in the bar graph (**Figure 3d**). We observe a reduction of approximately $9\ (\pm 3)$ nm in $\xi$ after two consecutive thermal cycles, scaling equivalently to the enhancement in the magnitude of $H_{EB}$ in one of the measured devices (Device 1). We find that the largest exchange bias is observed after the initial cooling of a freshly prepared device. While this might be related to $\xi$, one would need to carry out concurrent imaging and exchange bias measurements, which is beyond the scope of this work.

We further examined the influence of the vdW gap on the total energy of the $MnPS_3/Fe_3GeTe_2$ heterostructure using first-principles calculations (refer to **Figure S10**). Our results indicate a strict lower bound on the minimal vdW gap of 6 Å. Furthermore, we observed that the total energy variation for distances exceeding this limit is negligible. This suggests that the system remains stable for finite vdW gaps beyond the defined lower bound and can transition into metastable states with varying vdW gaps at minimal energy expense. The origin of such a significant variation in the inter-crystalline distance and the modified vdW gap can be attributed to the thermal expansion coefficient mismatch induced strain between $Fe_3GeTe_2$ and $MnPS_3$, which would be significant at the interface.[30] Despite the inherent limitations in the controllability of vdW gap alteration using this method, it is a straightforward and efficient approach that can be expanded to tune also other spintronic phenomena related to interfaces[31,32] within vdW heterostructures. However, our present capabilities limit the quantification of the structure and magnetic properties of the formed amorphous layers and provide a quantitative description of how they contribute to the evolution of exchange bias. Although several potential factors could contribute, such as strain-induced crystallization, amorphization, or chemical effects at the interface due to varying gap spacing, our current explanation has to remain phenomenological. However, we would like to stress that the amorphization and recrystallization induced by thermal cycling in van der Waals (vdW) heterostructure interfaces hold significant importance on its own. We recognize the need to address this issue using more sophisticated and sensitive techniques, which we plan to pursue in a subsequent study that goes beyond the scope of the current work.

## 3. Conclusion

In conclusion, we observed a huge EB in $MnPS_3$/$Fe_3GeTe_2$ vdW heterostructures despite $MnPS_3$ having a compensated collinear antiferromagnetic spin structure below the Néel temperature of 78 K. Notably, our experiments reveal that significant exchange bias appears only below 40 K, a phenomenon attributed to an inherent anomalous weak ferromagnetic moment emerging in $MnPS_3$. This phenomenon has remained largely unexplored thus far, and it originates from a small fraction of spins transitioning from an OOP Heisenberg-type order to an IP $XY$-type order. The magnitude of EB is found to exhibit an evolving nature that leads to a variation of $H_{EB}$ up to $1000\ \%$ achievable merely by thermal cycling. This evolving nature of EB observed is due to the modified interfacial registry of the vdW AFM/FM heterostructure characterized by a modified distance between the crystalline layers, as shown by cross-sectional

STEM measurements. High-resolution STEM measurements, combined with elemental analysis, reveal the migration of atoms across the vdW gap and the formation of an amorphous region on both FM and AFM layers at the interface. These findings emphasize the need for thorough interface characterization for instance by high-resolution imaging before drawing conclusions on the coupling at interfaces of magnetic van der Waals heterostructures. Our work thus challenges a widely spread belief that van der Waals materials form clean, well-defined, and defect-free interfaces in heterostructures. In particular, one has to note that chemical reactions at the interface do take place even if the materials are individually exhibiting well-defined surfaces. The observed remarkable EB in vdW AFM/FM heterostructures, subject to tunability through straightforward thermal cycling-induced vdW gap engineering reveals significant potential for the facile manipulation of EB in vdW heterostructures for magnetic storage and sensor applications. This simple strategy can be implemented for tuning other phenomena at interfaces within van der Waals heterostructures.

**4. Methods**

*$MnPS_3$/$Fe_3GeTe_2$ vdW heterostructure fabrication:* Bulk 'single crystals' of $MnPS_3$ and $Fe_3GeTe_2$ were obtained from our collaborators in Johannes Kepler University Linz and Max Planck Institute for Solid State Research Stuttgart, respectively. The bulk crystals were subjected to Raman spectroscopy measurements, and the obtained spectra agree well with the standard Raman modes expected for $MnPS_3$[33] and $Fe_3GeTe_2$[34] (refer to **Figure S11** in *Supporting information*). The bulk crystals were exfoliated down to a few layered flakes onto a Si/$SiO_2$ (300 nm) substrate inside a glovebox, with the $O_2$ and $H_2O$ levels maintained well below 0.5 ppm. A suitable few layered flakes of both $Fe_3GeTe_2$ and $MnPS_3$ were selected using an optical microscope. We employ a Polydimethylsiloxane (PDMS) and polymethylmethacrylate (PMMA) assisted dry transfer method inside the glovebox to transfer and stack multi-layered flakes of $Fe_3GeTe_2$ and $MnPS_3$ in the specified order. This results in the formation of an $MnPS_3/Fe_3GeTe_2$ vdW heterostructure with a clean interface. To minimize the risk of oxidation due to brief exposure of the heterostructure to air during wire-bonding of the voltage and current pads for anomalous Hall measurements, we cap the obtained heterostructure with an h-BN flake. Additionally, special care is taken to limit the air exposure time to below 30 minutes in all the devices measured before loading onto the high field cryostat sample space, where a low-pressure helium environment is maintained during this investigation unless otherwise stated. Atomic force microscopy is carried out to measure the thickness of $MnPS_3$ and $Fe_3GeTe_2$ flakes of the heterostructures (refer to **Figure S12** in *Supporting information*).

*Magneto-transport measurements:* The h-BN-capped $MnPS_3/Fe_3GeTe_2$ heterostructures were then wire-bonded and immediately loaded onto a variable temperature insert (VTI) cryostat in which we can apply a magnetic field up to 12 T for anomalous Hall effect (AHE) measurements. Since $MnPS_3$ is an insulating antiferromagnet, the current will only flow through the metallic $Fe_3GeTe_2$ layer. The cross-sectional area of the current channel along an $80\ \text{nm} \times 5\ \mu\text{m}$ $Fe_3GeTe_2$ flake (Device 1) is $4 \times 10^{-13}$ m$^2$, and is used to calculate the current density, which is approximately $2.5 \times 10^6$ Am$^{-2}$ for an applied current of $1\ \mu$A. A $1\ \mu$A current is transmitted through the $Fe_3GeTe_2$ flake along the $x$-direction and the anomalous Hall voltage ($V_{xy}$) is measured across the transverse terminals along the $y$-direction. The field is always applied in the OOP orientation for the field-cooling process and the field sweeping. A Keithley 2400 source meter is used to flow current through the device, and a Keithley 2182a nanovoltmeter is used to measure voltage.

*DC SQUID Magnetometry Measurements:* Magnetic hysteresis loops ($M$ vs. $H$) and temperature-dependent magnetization curves ($M$ vs. $T$) were acquired for a bulk $MnPS_3$ crystal using a Quantum Design SQUID MPMS3 Magnetometer. The $M$ vs. $H$ measurements involved sweeping the magnetic field in both IP and OOP orientations relative to the crystal's bulk structure. Meanwhile, *M* vs. *T* measurements were conducted over a temperature range of 5 K to 100 K, applying a magnetic field of 50 Oe in an OOP orientation.

*Scanning nitrogen-vacancy center magnetometry (S-NVM) measurements:* S-NVM is a potent, non-invasive imaging technique suitable for probing weak magnetic fields on the order of μT with a spatial resolution of (30-100 nm), especially suitable for the investigation of magnetism in vdW nanomagnets.[35 – 37] An all-diamond atomic force tip containing a single NV near the apex of the tip (≈10-20 nm) is brought close to the sample's surface.[38] The NV center hosts an optically addressable spin that is initialized and read out all optically while it is coherently manipulated by externally applied microwaves. The upper spin levels show a Zeeman effect, which allows for a precise measurement of the stray magnetic field along the NV quantization axis.[39] S-NVM measurements are performed in a commercial cryogenic atomic force microscopy system (AttoCube, AttoLiquid 1000) with a home-built NV magnetometry setup (details in Ref [40]). For AFM operation, the NV tip is mounted on a quartz capillary glued to a quartz tuning fork. The tuning fork is operated in shear mode with amplitude feedback to keep the tip at a constant height above the surface. The tip is positioned by Piezo positioners and scanners, providing a scan range of 15 μm × 15 μm. The complete atomic force microscopy setup is placed in a sample chamber at 100 mbar He atmosphere located in the liquid bath cryostat at a base temperature of 4.3 K. A 0.5 T vector magnet is used to apply an external magnetic field. The emission of a 532 nm laser (Laser Quantum GEM) is collimated on a 0.81 NA objective (AttoCube LT-APO) and subsequently focused on the NV. The resulting photoluminescence (PL) is collected via the excitation path and separated by a dichroic mirror (Thorlabs DMLP567). The PL photons are collected by an avalanche photodiode (Excelitas SPCM-AQRH-33). The microwave signal for coherent manipulation of the NV spin is produced by a signal generator (SRS SG384) and an IQ modulator (Polyphase Microwave AM0350A), amplified (Minicircuit ZHL-42W+), and transmitted by a bond wire~ 100 nm close to the NV tip. An arbitrary waveform generator (Spectrum instrumentation DN2.663-4) is used to pulse the microwave driving and the laser emission. The resonance frequency of the spin transition is recorded at each pixel to determine the corresponding stray magnetic field from the Zeeman-induced frequency shift. The NV is excited with a 500 μs laser pulse, which initializes the NV into the $m_s = 0$ spin state. Next, a near-resonant microwave $\pi$ pulse is applied to the NV center. The resulting spin state is read out by a second 500 $\mu s$ laser pulse that also reinitializes the spin state for the following pulse cycle. Two different measurement schemes are used: a) full optical detected magnetic resonance spectra are recorded at each pixel, and b) a feedback protocol is used to track the resonance frequency of the NV. While the first scheme requires more integration time but is less prone to measurement artifacts, the second scheme provides higher field sensitivity, resulting in a lower noise level. S-NVM is carried out at a base temperature of 4.5 K, with an external bias field of 150 mT applied along the determined NV axis before the measurements. The NV axis of the tip used for this work was measured to be 56° (±2°) with the OOP direction. This pre-application of an external bias field aims to optimize sensor performance at low temperatures.[41]

*Scanning transmission electron microscopy (STEM) Measurements:* The STEM imaging is done using a double-corrected TFS Spectra 300 microscope operated in STEM mode at 300 kV. The microscope is equipped with a high-brightness X-FEG, monochromated source, and a piezo CompuStage, and can achieve sub-Ångström resolution. The transmission electron microscopy (TEM) lamella is tilted, so the $Fe_3GeTe_2$ is in the <0001> zone axis. A high-angle annular dark-field (HAADF) detector is used to image the interface between the $Fe_3GeTe_2$ and $MnPS_3$. The intercrystalline distance between the $Fe_3GeTe_2$ and $MnPS_3$ stacks responsible for the exchange coupling is measured using high-resolution STEM. The required TEM lamella is prepared using the focused ion beam (FIB) technique. The lamella is prepared by cutting a stack of $Fe_3GeTe_2$/$MnPS_3$ on a (300 nm) $SiO_2$/Si wafer and encapsulated in h-BN using a FEI make Helios 460F dual beam FIB-SEM. A 5 nm thick Pd layer is deposited on top of the stack to prevent charging. Two protective layers of electron beam deposited Pt and ion beam Carbon were deposited atop the area of interest to reduce the sample contamination due to gallium implantation. A TEM lamella with an electron transparent window of ~ 50 nm thickness is cut from the chip, and the distance between the $Fe_3GeTe_2$ and $MnPS_3$ layers is measured (see **Figure S7** in *Supporting information* for a summary of stepwise STEM cross-section lamella fabrication). To study the change in the inter-crystalline distance with cooling cycles, it is preferred to cool down the entire crystal rather than the TEM lamella since the process of lamella fabrication reduces the sample conditions due to the strains introduced due to the production of thin TEM lamella and Ga implantation. The $SiO_2$/Si chip consisting of the vdW heterostructure is cooled down to 77 K by dropping the chip into $LN_2$ in a dewar and waiting at least 8 hours until all the $LN_2$ evaporates. After the chip warms up to room temperature, it is dried using an $N_2$ gun to remove any moisture condensing on the surface. Additional TEM lamellae were prepared after each cooling cycle to measure the inter-crystalline distance. Energy dispersive X-ray (EDX) spectroscopy is carried out using an FEI Titan TEM equipped with a Schottky field emission gun operated at 200 kV, a CEOS probe aberration corrector, a high angle annular dark-field detector (HAADF) and a Super-X EDX detection system. Elemental maps and profiles were obtained using the Thermo Scientific Velox software. The results are summarized in **Figure S9** in *Supporting information*.

*First-principles calculations:* First-principles calculations were performed using the VASP software package [42-45]. Distances between 5 and 12 Å were chosen and a full spin-polarized self-consistency with FM and AFM starting ground states were chosen for FGT and MNPS3 respectively. The calculations were performed using the GGA exchange correlation functional[46] and converged on a 8 by 8 k-mesh with an energy cutoff of 270eV. The results are summarized in **Figure S10** in *Supporting information*.

**Acknowledgments**

APB and AK contributed equally to this work. We acknowledge funding from Alexander von Humboldt Foundation for Humboldt Postdoctoral Fellowship (Grant number: Ref 3.5-IND-1216986-HFST-P), EU Marie-Curie Postdoctoral Fellowship ExBiaVdW (Grand Id: 101068014), Deutsche Forschungsgemeinschaft (DFG, German Research Foundation) – Spin + X TRR 173–268565370 (Projects No. A01, No. B02 and A12), DFG Project No. 358671374, Graduate School of Excellence Materials Science in Mainz (MAINZ) GSC 266, the MaHoJeRo (DAAD Spintronics network, Projects No. 57334897 and No. 57524834), the Research Council of Norway (Centre for Quantum Spintronics - QuSpin No. 262633), The European Union's Horizon 2020 Research and Innovation Programme under grant agreement 856538 (project "3D MAGIC"). S.F. acknowledges the China Scholarship Council for financial support. Additionally, we wish to express our gratitude to Dr. Arnab Bose for the insightful discussions.